\documentclass[10pt]{IEEEtran}
\usepackage[T1]{fontenc}
\usepackage{textcomp}
\usepackage{txfonts}

\usepackage[utf8]{inputenc}
\usepackage[english]{babel}
\usepackage{csquotes}
\usepackage[pdftex,hidelinks]{hyperref}

\usepackage{cite}
\usepackage{graphicx}
\usepackage{tikz}
\usetikzlibrary{calc}

\usepackage[detect-all]{siunitx}
\usepackage[capitalise]{cleveref}
\crefname{figure}{Fig.}{Fig.}
\Crefname{figure}{Fig.}{Fig.}

\usepackage{microtype}

\newcommand*{\xr}[1]{\ensuremath{_{#1}}}

\begin{document}

\bstctlcite{BSTcontrol}%

\title{Beamline Spectroscopy of Integrated Circuits With Hard X-ray Transition Edge Sensors at the Advanced Photon Source}%

\author{T.~Guruswamy, L.~Gades, A.~Miceli, U.~Patel, O.~Quaranta%
\thanks{Manuscript received DDD; revised DDD: accepted DDD.
This research was supported by Laboratory Directed Research and Development (LDRD) funding from Argonne National Laboratory, 
and the Accelerator and Detector R\&D program in Basic Energy Sciences' Scientific User Facilities (SUF) 
Division at the Department of Energy; used resources of the Advanced Photon 
Source and Center for Nanoscale Materials, U.S. Department of Energy Office of 
Science User Facilities operated for the DOE Office of Science by 
Argonne National Laboratory under Contract No. DE-AC02-06CH11357, and the Pritzker Nanofabrication Facility of the Institute for Molecular 
Engineering at the University of Chicago, which receives support from Soft and 
Hybrid Nanotechnology Experimental (SHyNE) Resource (NSF ECCS-2025633), a node of 
the National Science Foundation's National Nanotechnology Coordinated 
Infrastructure. \textit{(Corresponding author: T. Guruswamy.)}}%
\thanks{The authors are with the X-ray Science Division, Argonne National Laboratory, Lemont, IL 60439 USA (e-mail: tguruswamy@anl.gov).}%
}
\maketitle

\begin{abstract}
At Argonne National Laboratory, we are developing hard X-ray (2 to 20 keV)
Transition Edge Sensor (TES) arrays for beamline science. The significantly
improved energy resolution provided by superconducting detectors compared to
semiconductor-based energy-dispersive detectors, but with better
collection efficiency than wavelength-dispersive instruments, will enable
greatly improved X-ray emission and absorption spectroscopic
measurements. A prototype instrument with 24 microwave-frequency multiplexed
pixels is now in testing at the Advanced Photon Source (APS) 1-BM beamline.
Initial measurements show an energy resolution ten times better (\SI{150}{eV} compared to $<$ \SI{15}{eV}) than the
silicon-drift detectors currently available to APS beamline users, and
in particular demonstrate the ability to resolve closely-spaced emission lines in
samples containing multiple transition metal elements, such as integrated
circuits. Comparing fluorescence spectra of integrated circuits measured with our TESs at 
the beamline to those measured with silicon detectors, we find emission lines and elements
largely hidden (e.g. Hf alongside Cu) from a semiconductor-based detector but well resolved by a TES.
This directly shows the strengths of TES-based instruments in fluorescence mapping.
\end{abstract}

 \section{Introduction}
 \IEEEPARstart{S}{uperconducting} photon detectors offer an order of magnitude improvement in energy resolution over semiconductor-based detectors.
 This has enabled them to make a significant impact in a variety of scientific fields, including astronomy, particle physics, and X-ray science~\cite{Irwin2005,Ullom2015}.
 Instruments based on arrays of superconducting Transition Edge Sensors (TESs) are now being deployed at a number of X-ray light sources worldwide~\cite{Doriese2017,Lee2019}.
 At the Advanced Photon Source (APS), Argonne National Laboratory, we are developing a hard X-ray TES array-based instrument for X-ray Fluorescence (XRF) and X-ray Absorption (XAS) spectroscopy.
 Once commissioned, this instrument will be made available for beamline user science.
 
 With their high photon collection efficiency and excellent energy resolution, one potential application for TES-based X-ray spectrometers is high spatial resolution X-ray fluorescence mapping.
 In this technique, an image is created by scanning the sample with a very focused X-ray beam, measuring X-ray fluorescence spectra at each sampled spatial position to obtain relative elemental abundances.
 Collecting photons with a high efficiency is important to map quickly, and a high energy resolution is important to correctly identify the fluorescence lines and determine the elemental ratios at each sample point.
 TESs have an advantage over narrow band, low collection efficiency wavelength-dispersive instruments by covering a wide energy range in a single measurement, and for radiation-sensitive samples which do not allow detector inefficiencies to be compensated for by increasing the incident beam brightness.
 They also have a large advantage over lower-resolution semiconductor-based energy dispersive instruments for complex samples with mixed $3d$, $4d$, and $5d$ transition metal elements with overlapping $K$ and $L$-shell emission lines.
 Integrated circuits are an example of samples in this latter category.
 Besides their semiconductor substrate, integrated circuits consist of complex 3-D geometries of metals and metal compounds.
 Precisely mapping these features non-destructively requires X-ray wavelengths due to their size~\cite{Levine1998,Xu2003}.
 Such a map might be needed for independent verification of correct manufacture, reverse engineering, or failure analysis~\cite{Mahmood2015}.
 
 The applications of a TES XRF mapping instrument at a synchrotron are similar to those of TES imaging spectrometers in electron microscopes~\cite{Ullom2015} and particle-induced X-ray emission (PIXE) instruments~\cite{Palosaari2016}.
 A synchrotron-based instrument can additionally make use of the monochromatic, very high photon flux X-ray beam tunable over a wide energy range (\SIrange{3}{100}{keV} at the APS),
 nano-scale control of the spatial and temporal distributions of incident X-rays, and exotic (e.g. high pressure or temperature) sample environments.
 This allows for measurements of very dilute or weakly scattering samples, as well as simultaneous measurements with other techniques such as ptychography, Energy-dispersive Diffraction (EDD), Compton imaging, Extended X-ray Absorption Fine Structure (EXAFS), and Resonant Inelastic X-ray Scattering (RIXS).
 Though our current APS instrument with its prototype TES arrays is not intended for imaging applications, as a demonstration of the concept we have used it to collect fluorescence spectra
 from three different integrated circuit chips at the APS beamline, and made a direct comparison against spectra of the same samples collected with a silicon drift detector.
 This is both a characterization of our system and a simple demonstration of what future purpose-built TES XRF mapping instruments~\cite{Weichman2020} might be able to achieve.
 
 \section{Instrument Design}
 
 Our standard TES pixel design features \SI{120}{\micro{}m}-square MoCu bilayers connected to ``sidecar''-style absorbers~\cite{Guruswamy2020,Patel2020}.
 Normal metal (Cu) features including banks and bars atop the bilayer act to reduce noise~\cite{Ullom2004}.
 The thin SiN membrane supporting each pixel thermally isolates it from the substrate; the thermal conductance is further reduced by a series of perforations around its perimeter.
 Here the absorber consists of a \SI{700}{\micro{}m}-square of \SI{1}{\micro{}m} thick sputtered Au, which is sufficient to stop 14\% of \SI{20}{keV} photons.
 Devices with an additional \SI{10}{\micro{}m} electroplated Bi on the absorbers are now being fabricated; this will raise the stopping power at \SI{20}{keV} to 63\%.
 As Bi has a heat capacity $C$ at least an order of magnitude smaller than Au, we do not expect the energy resolution ($\propto C^{1/2}$) to be significantly degraded~\cite{Brown2008}.
 We have previously shown adding Bi to similar TES arrays results in no measurable change in energy resolution, and the larger grain size of electroplated Bi is
 preferred over evaporated Bi to avoid energy trapping at grain boundaries~\cite{Yan2017a}.
 
 We have fabricated and tested multiple 24-pixel arrays of this design at Argonne National Laboratory.
 Detailed characterization measurements are consistent with those previously reported~\cite{Guruswamy2020}; our typical bilayers have a superconducting transition temperature near \SI{87.5}{mK}, and a normal state resistance near \SI{8}{m\ohm}.
 The thermal conductance $G$ and heat capacity $C$ are largely controlled by the absorber dimensions; we find $G \sim \SI{500}{pW.K^{-1}}$ and $C \sim \SI{4}{pJ.K^{-1}}$. 
 Si collimators with etched \SI{400}{\micro{}m}-square apertures and \SI{10}{\micro{}m} thick electroplated Au are mounted over the TES array to ensure photons are incident only on the absorbers.
 The collimators include lithographically-defined step features at their corners, which fix both their horizontal position and vertical separation relative to the TES array.
 \Cref{fig:diagrams}(b) is a photograph of two 24-pixel arrays mounted with collimators.
 The collimator edge has cutouts to allow wire bonds and Cu spring clamps access to the TES array underneath. Al wire bonds connect to the readout circuitry,
 and Au bonds to the perimeter of the TES array and collimator aid thermalization.
 
 The TES array readout is multiplexed using microwave frequency multiplexing~\cite{Mates2017}.
 Each pixel is coupled to an RF SQUID and superconducting microwave resonator (bandwidth \SI{300}{kHz}) on a nearby multiplexing chip fabricated by the Quantum Sensors Group at NIST.
 Appropriate shunt resistors (\SI{300}{m\ohm}) and fixed inductors (\SI{100}{nH}) are also included in the TES bias circuit.
 The TESs are voltage biased in series, and the SQUID responses are linearized using flux ramp modulation~\cite{Mates2012}, so in total only one coaxial pair and two DC pairs of wires are needed to read out an entire array.
 The generation, measurement, and demodulation of the appropriate tones for each resonator is done using a ROACH2-based FPGA system~\cite{Gard2018}.
 In its current configuration, a single ROACH2-based readout rack of ours provides continuous digitized readout of up to 128 TES pixels at a \SI{62.5}{kHz} sampling rate.
 Raw data from each pixel are then processed into pulse records using software-based triggering, and saved.
 Offline optimal filtering with gain, arrival time and time drift correction is used to extract the best estimate of pulse energy from each saved record~\cite{Fowler2016}.
 
 \section{Measurement Methods}
\begin{figure}
  \centering
  \small
  \begin{tikzpicture}
   \coordinate (a) at (-2,1);
   \coordinate (b) at (-1,0);
   \draw[gray, ultra thick] (a) -- ++(1,0) -- ++(0,-2) -- ++(-1,0);
   \draw[gray, very thick] (a) ++(0,-0.2) -- ++(0.8,0) -- ++(0,-1.6) -- ++(-0.8,0);
   \draw[blue, very thick] (a) ++(0,-0.4) -- ++(0.6,0) -- ++(0,-1.2) -- ++(-0.6,0);
   
   \draw[yellow, ultra thick] (a) ++(0.2,-0.6) rectangle ++(0.2, -0.8);
   \draw[yellow, fill=yellow, line width=4] (a) ++(0,-1) -- ++(0.2,0);
   
   \draw[red, fill=red] (-0.3,2.4) rectangle (0.3,2);
   \draw[black, ultra thick] (-0.5,2) -- (0.5,2);
   \draw[red, ultra thick] (0,2.1) -- (0,0);
   \draw[red, very thick] (0,0) -- ++(-1,0);
   \draw[violet, line width=4] (-0.2,-0.2) -- ++(0.4,0.4);
   \draw[orange] (-0.45,-0.55) -- ++(1,1);
   
   \draw[white, line width=2] (b) ++(0,0.2) -- ++(0,-0.4);
   \draw[white, line width=2] (b) ++(-0.2,0.2) -- ++(0,-0.4);
   \draw[white, line width=2] (b) ++(-0.4,0.2) -- ++(0,-0.4);
   \draw[white, line width=2] (b) ++(-0.6,0.15) -- ++(0,-0.3);
   
   \draw[gray, thin] (b) ++(0,0.2) -- ++(0,-0.4);
   \draw[gray, thin] (b) ++(-0.2,0.2) -- ++(0,-0.4);
   \draw[gray, thin] (b) ++(-0.4,0.2) -- ++(0,-0.4);
   \draw[gray, thin] (b) ++(-0.6,0.15) -- ++(0,-0.3);
   
   \draw[green, line width=2] (b) ++(-0.75,0.2) -- ++(0,-0.4);
   
   \draw[|-|] (b) ++(-0.7,-1.4) -- ++(0.7,0) node[label={right:\SI{90}{mm}}] {};
   \draw[|-|] (-1,-1.8) -- ++(1,0) node[label={right:\SI{115}{mm}}] {};
   \node[anchor=north west,fill=white] at ($(a)+(0,1.25)$) {(a)};
   
   \node[anchor=north west] at (1.5,2.6) {\includegraphics[width=4.2cm]{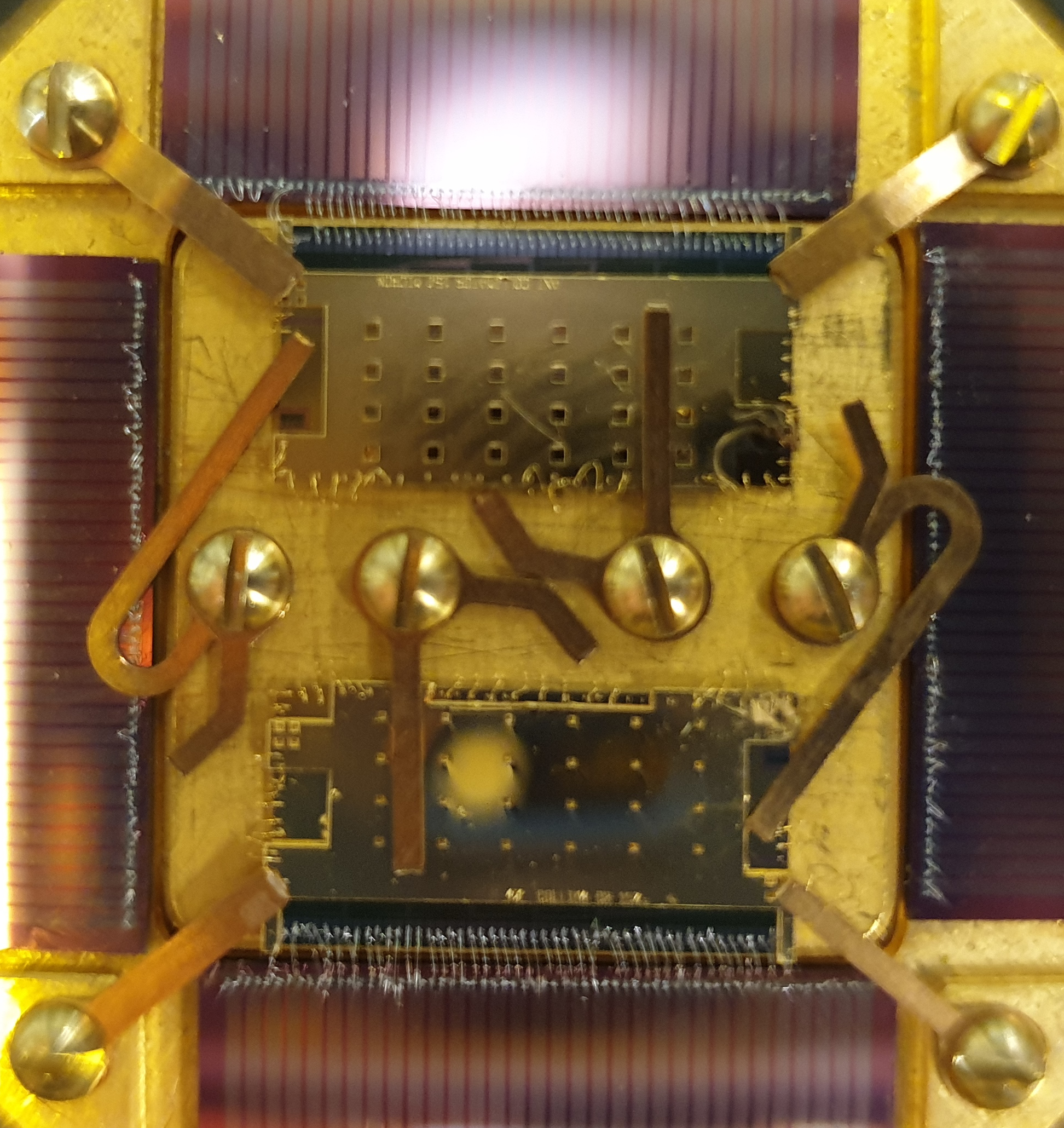}};
   \node[anchor=north east,fill=white,fill opacity=0.75,text opacity=1] at ($(a)+(4.5,1.25)$) {(b)};
  \end{tikzpicture}
  
  \caption{(a) Schematic of experimental setup, top-down view (not to scale). Monochromatic X-ray beam from synchrotron (red) is reduced in size by upstream slits (black).
  Sample (violet) is mounted on Kapton tape in a standard scattering geometry, at a \SI{45}{\degree} orientation to both beam and detector.
  Scattered X-rays pass through windows in the cryostat outer vacuum jacket and inner radiation shields, mu-metal magnetic shielding, and detector box (yellow), illuminating the TES array (green).
  (b) Photograph of two 24-pixel TES arrays, with collimators, mounted in detector box. The X-rays illuminate approximately the entire center square.
  Measurements reported here are from the top device with \SI{400}{\micro{}m} square collimator apertures; bottom collimator has \SI{150}{\micro{}m} square apertures.
  See \cite[Fig. 1]{Guruswamy2020} for images of the TES array itself.
  \label{fig:diagrams}}
 \end{figure}
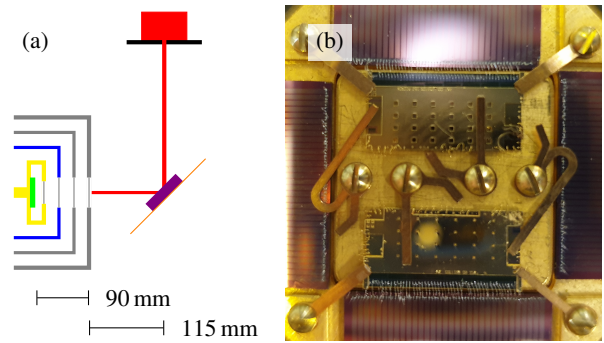

 A 24-pixel TES die was mounted in our combined He$_3$ adsorption/ADR cryostat, which has been installed at the APS 1-BM-C beamline experimental station since November 2019.
 This cryostat has a ``snout'' which allows for the TES sensor array to be oriented towards the sample, in the horizontal plane of the beam.
 It maintains the array at its operating temperature of \SI{60}{mK} for up to \SI{80}{hours}, allowing for efficient use of X-ray beam time.
 X-rays are allowed through the cryostat vacuum jacket by a circular Beryllium window, then through the internal radiation shields,
 mu-metal passive magnetic shielding, and detector housing by windows of Al foil as shown in \cref{fig:diagrams}(a). The outermost window \SI{90}{mm} from the sensors with radius \SI{20}{mm} limits the instrument field of view.
 
 Initial beamline characterization of the TES pixels, by detecting fluorescence from simple metal foils illuminated by a monochromatic X-ray beam at an energy of at least $\SI{15}{keV}$, demonstrated a typical energy resolution of \SIrange{12}{15}{eV}.
 This was determined by fits to the K\xr{\alpha} emission line complexes of various 3$d$ metals (Mn, Fe, Cu), accounting for their intrinsic line width.
 The best energy resolutions were obtained when the incident photon rates were kept below \SI{3}{Hz} per pixel and the average measured photon energy kept below \SI{12}{keV}.
 Increasing either the photon energy or photon rate significantly beyond this worsened the energy resolution, despite the pixels remaining far from saturation; we are investigating whether this degradation is related to thermal or electrical cross-talk between pixels, or pulse pile-up on the same pixel.
 
  \begin{figure*}
  \centering
  \includegraphics[width=\textwidth]{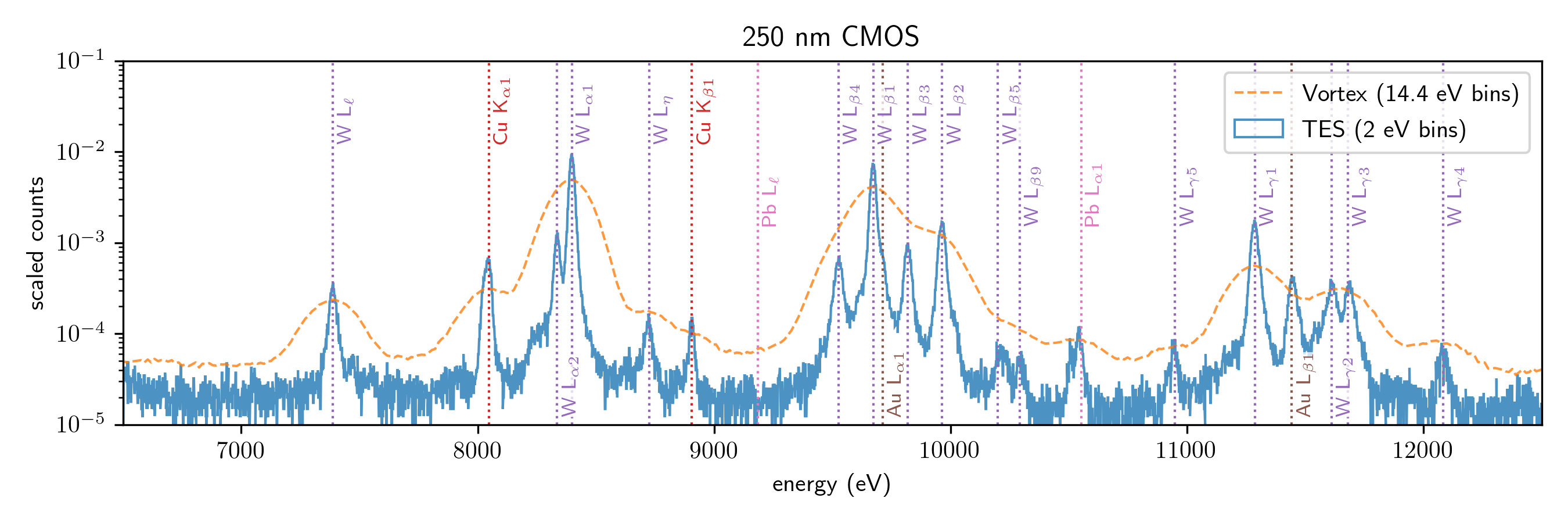}
  \caption{Fluorescence spectra of a \SI{250}{nm} CMOS integrated circuit chip, measured with a TES sensor (blue solid line) and a Vortex silicon-drift detector (orange dashed line).
  Emission lines from W and Cu are visible in both measurements; prominent peaks are labeled with their corresponding element and line name.\label{fig:250nm}}
 \end{figure*}
 \begin{figure*}
  \centering
  \includegraphics[width=\textwidth]{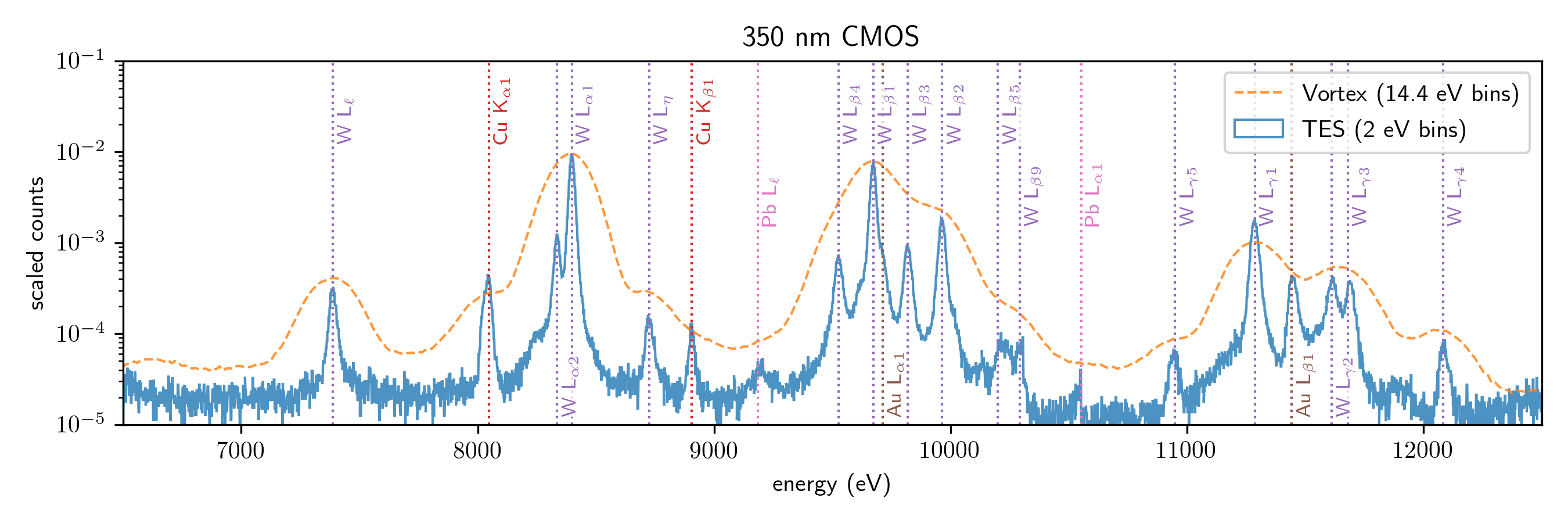}
  \caption{Fluorescence spectra of a \SI{350}{nm} CMOS integrated circuit chip, measured with a TES sensor (blue solid line) and a Vortex silicon-drift detector (orange dashed line).
  Prominent peaks are labeled with their corresponding element and line name.\label{fig:350nm}}
 \end{figure*}
   \begin{figure*}
  \centering
  \includegraphics[width=\textwidth]{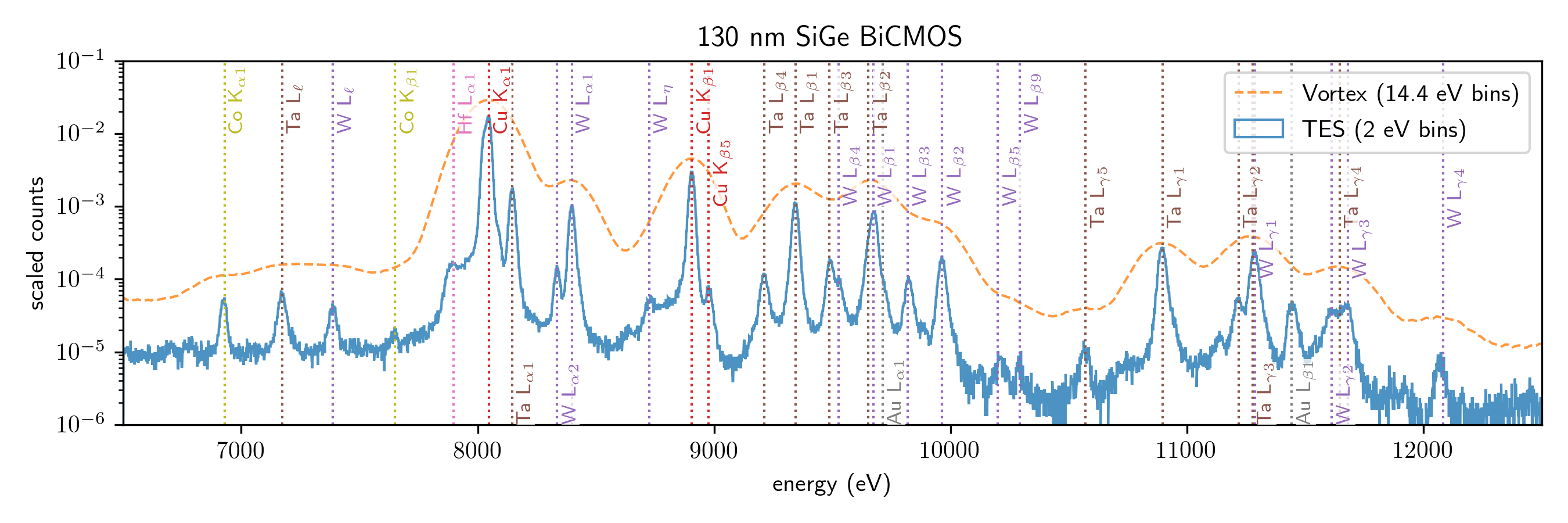}
  \caption{Fluorescence spectra of a \SI{130}{nm} SiGe BiCMOS integrated circuit chip, measured with a TES sensor (blue solid line) and a Vortex silicon-drift detector (orange dashed line).
  Relative to the other chips, emission lines from Co, Ta, and Hf are also present, but are clearly resolved only in the TES-measured spectrum.\label{fig:faspax}}
 \end{figure*}
 
 Three integrated circuit chips, unpackaged bare dies, were selected for the proof-of-concept XRF measurements.
 These samples were all of different process technologies, manufactured at commercial foundries.
 We chose a \SI{250}{nm} CMOS chip, a \SI{350}{nm} CMOS chip, and a \SI{130}{nm} SiGe BiCMOS chip, all approximately \SI{10}{mm} by \SI{20}{mm}.
 The exact metal stacks used are proprietary formulations, so the elemental compositions of these chips were not known to us beforehand.
 The chips were secured to Kapton film, and held with the front face of the chip exposed in the standard \SI{45}{\degree} scattering geometry shown in \cref{fig:diagrams}(a).
 The \SI{22.5}{keV} monochromatic X-ray beam illuminated a \SI{5}{mm} by \SI{3}{mm} area near the center of each chip.
 This beam spot size was chosen to ensure a low (less than \SI{2}{Hz} per TES pixel) incident count rate at the detector.
 Spectra were collected with the TES array for at least 8 hours, and then with a single-pixel Vortex silicon-drift detector for a similar length of time (at a much higher count rate, above $\SI{1}{kHz}$).
 
 The Vortex silicon-drift detector was separately calibrated using fluorescence from the same Mn, Fe, and Cu metal foils.
 The TES pixels were calibrated directly from the integrated circuit spectra after the prominent peaks in the measured data were identified.
 For each pixel and each measurement, a power-law calibration curve was fitted to the identified peaks in the optimal-filtered pulse height histogram.
 We found the response of each pixel was linear over the energy range \SIrange{4}{20}{keV}.
 In this series of measurements, 15 of the 24 TES pixels in the array both measured sufficient counts and calibrated correctly.
 The inoperative pixels are primarily attributed to wiring defects or post-fabrication damage.
 The calibrated data from each working pixel were co-added and binned to produce an energy histogram for each sample, each containing \numrange{e5}{e6} total counts.
 
 \section{Results and Discussion}

 The measured spectra are presented in \cref{fig:250nm,fig:350nm,fig:faspax}.
 Counts from the TES pixels and the Vortex have been normalized by the total counts collected for easier comparison.
 As expected, the measurements expose the presence of several transition metals.
 Emission lines from W and Cu dominate the \SI{250}{nm} CMOS and \SI{350}{nm} CMOS spectra (\cref{fig:250nm,fig:350nm}).
 Most peaks are well resolved by the TES, except for Au L\xr{\alpha1} (\SI{9713}{eV}) which is visible only as a high-energy shoulder of the W L\xr{\beta1} (\SI{9672}{eV}) peak.
 A very small number of background counts are introduced to the TES-measured spectra by fluorescence from the Au collimator and absorber, and stray fluorescence from Pb elsewhere in the beamline hutch.
 These contributions were verified by comparison to the XRF spectra of a bare Si wafer.
 Otherwise, there are very few background counts and little evidence for low-energy tails or non-Gaussian peak shapes, resulting in easily interpretable spectra.
 W and Cu are known to be present in integrated circuit interconnects and vias~\cite{Levine1998a}.
 The \SI{250}{nm} CMOS and \SI{350}{nm} CMOS spectra are only distinguished by a small difference in the ratios of Cu to W; the \SI{250}{nm} chip has \num{1.5} times more Cu than the \SI{350}{nm} chip, measured relative to W.
 This is consistent for both the TES and Vortex-measured spectra.
 
 \Cref{fig:faspax} shows the measured fluorescence spectra from the \SI{130}{nm} SiGe BiCMOS chip.
 Although both the TES-measured and Vortex-measured spectra show clear differences from the CMOS chips, in particular a significant increase in Cu, only in the TES-measured spectra is it straightforward to identify the additional presence of Co, Ta and Hf.
 In particular, the primary lines from these elements -- Hf L\xr{\alpha} (\SI{7899}{eV}) and Ta L\xr{\alpha} (\SI{8146}{eV}) -- cannot be separated from the Cu K\xr{\alpha} peak (\SI{8048}{eV}) by the Vortex.
 Quantitative analysis of the TES-measured spectra can therefore be performed much more simply and with much more statistical confidence.
 In SiGe processes, Co silicides are often used to form the connections between semiconductor elements and the metallic interconnects~\cite{Cressler2018}, TaN is a common resistor material, and Hf may be a component of high-$k$ dielectrics for capacitors.
 We also know that as process nodes have reduced in feature size, Cu has become preferred over other metals for interconnects due to its lower resistivity~\cite{Bohr1995}.
 
 The long collection time of the TES-measured spectrum can be easily shortened without impacting the total counts collected by scaling up the number of pixels, at least to our readout capabilities (128 pixels).
 Further increases may be realized by modifying the pixel design to reduce the pulse decay time, in particular by increasing the thermal conductance. This has only a very weak effect on the energy resolution.
 
 \section{Conclusions}
 The spectra collected demonstrate that a hard X-ray TES-based energy dispersive detector at a synchrotron can provide high-resolution, broadband spectroscopy to efficiently characterize samples with complex combinations of transition metals, like integrated circuits.
 The current energy resolution of our pixels, \SIrange{12}{15}{eV}, is already sufficient to demonstrate large advantages over a Vortex silicon-drift detector (energy resolution at best $\SI{130}{eV}$).
 Weak emission lines within one hundred eV in energy of a prominent peak are completely invisible to the Vortex, but are able to be identified by the TES array.
 Although we cannot confirm the exact elemental abundances in these samples due to their proprietary nature, we can claim success in distinguishing different process nodes.
 While our current instrument does not have imaging capabilities, future instruments will do so~\cite{Weichman2020}; and these, possibly in combination with highly focused light sources like those in development as part of the ongoing APS Upgrade (APS-U), will be able to perform detailed fluorescence mapping to non-destructively analyze integrated circuits at high spatial resolutions.
 As integrated circuit design and fabrication processes continue to become more complex and separated from their end users, this capability to map their internal structures will become increasingly essential, and we believe TES-based X-ray instruments will play a major part.
 
 \section*{Acknowledgment}
 \addcontentsline{toc}{section}{Acknowledgment}
 The authors would like to thank M. Wojcik for assistance with the beamline setup, and the members of the Quantum Sensors Group, NIST (Boulder, CO USA), for their advice on TES design as well as providing and helping implement the $\mu$-mux readout system.
 
 \interlinepenalty=10000
 \bibliography{asc20_manuscript,bibtex/ieeecfg}

\newcommand{\noopsort}[1]{}
\begin{thebibliography}{10}
\providecommand{\url}[1]{#1}
\providecommand{\doi}[1]{doi: \nolinkurl{#1}}
\csname url@samestyle\endcsname
\providecommand{\newblock}{\relax}
\providecommand{\bibinfo}[2]{#2}
\providecommand{\BIBentrySTDinterwordspacing}{\spaceskip=0pt\relax}
\providecommand{\BIBentryALTinterwordstretchfactor}{4}
\providecommand{\BIBentryALTinterwordspacing}{\spaceskip=\fontdimen2\font plus
\BIBentryALTinterwordstretchfactor\fontdimen3\font minus
  \fontdimen4\font\relax}
\providecommand{\BIBforeignlanguage}[2]{{%
\expandafter\ifx\csname l@#1\endcsname\relax
\typeout{** WARNING: IEEEtran.bst: No hyphenation pattern has been}%
\typeout{** loaded for the language `#1'. Using the pattern for}%
\typeout{** the default language instead.}%
\else
\language=\csname l@#1\endcsname
\fi
#2}}
\providecommand{\BIBdecl}{\relax}
\BIBdecl

\bibitem{Irwin2005}
K.~Irwin and G.~Hilton, ``Transition-edge sensors,'' in \emph{Cryogenic
  Particle Detection}.\hskip 1em plus 0.5em minus 0.4em\relax {Springer,
  Berlin, Heidelberg}, 2005, pp. 63--150.
  \href{http://dx.doi.org/10.1007/10933596_3}{\doi{10.1007/10933596_3}}

\bibitem{Ullom2015}
J.~N. Ullom and D.~A. Bennett, ``Review of superconducting transition-edge
  sensors for x-ray and gamma-ray spectroscopy,'' \emph{Supercond. Sci.
  Technol.}, vol.~28, no.~8, p. 084003, Aug. 2015.
  \href{http://dx.doi.org/10.1088/0953-2048/28/8/084003}{\doi{10.1088/0953-2048/28/8/084003}}

\bibitem{Doriese2017}
W.~B. Doriese \emph{et~al.}, ``A practical superconducting-microcalorimeter
  {{X}}-ray spectrometer for beamline and laboratory science,'' \emph{Rev. Sci.
  Instrum.}, vol.~88, no.~5, p. 053108, May 2017.
  \href{http://dx.doi.org/10.1063/1.4983316}{\doi{10.1063/1.4983316}}

\bibitem{Lee2019}
S.-J. Lee \emph{et~al.}, ``Soft {{X}}-ray spectroscopy with transition-edge
  sensors at {{Stanford Synchrotron Radiation Lightsource}} beamline 10-1,''
  \emph{Rev. Sci. Instrum.}, vol.~90, no.~11, p. 113101, Nov. 2019.
  \href{http://dx.doi.org/10.1063/1.5119155}{\doi{10.1063/1.5119155}}

\bibitem{Levine1998}
Z.~H. Levine, A.~R. Kalukin, S.~P. Frigo, I.~McNulty, and M.~Kuhn,
  ``Tomographic reconstruction of an integrated circuit interconnect,''
  \emph{Appl. Phys. Lett.}, vol.~74, no.~1, pp. 150--152, Dec. 1998.
  \href{http://dx.doi.org/10.1063/1.123135}{\doi{10.1063/1.123135}}

\bibitem{Xu2003}
G.~Xu, D.~E. Eastman, B.~Lai, Z.~Cai, I.~McNulty, S.~Frigo, I.~C. Noyan, and
  C.~K. Hu, ``Nanometer precision metrology of submicron {{Cu}}/{{SiO2}}
  interconnects using fluorescence and transmission x-ray microscopy,''
  \emph{J. Appl. Phys.}, vol.~94, no.~9, pp. 6040--6049, Oct. 2003.
  \href{http://dx.doi.org/10.1063/1.1614430}{\doi{10.1063/1.1614430}}

\bibitem{Mahmood2015}
K.~Mahmood, P.~L. Carmona, S.~Shahbazmohamadi, F.~Pla, and B.~Javidi,
  ``\BIBforeignlanguage{EN}{Real-time automated counterfeit integrated circuit
  detection using x-ray microscopy},'' \emph{\BIBforeignlanguage{EN}{Appl.
  Opt.}}, vol.~54, no.~13, pp. D25--D32, May 2015.
  \href{http://dx.doi.org/10.1364/AO.54.000D25}{\doi{10.1364/AO.54.000D25}}

\bibitem{Palosaari2016}
M.~R.~J. Palosaari \emph{et~al.}, ``Broadband {{Ultrahigh}}-{{Resolution
  Spectroscopy}} of {{Particle}}-{{Induced X Rays}}: {{Extending}} the
  {{Limits}} of {{Nondestructive Analysis}},'' \emph{Phys. Rev. Applied},
  vol.~6, no.~2, p. 024002, Aug. 2016.
  \href{http://dx.doi.org/10.1103/PhysRevApplied.6.024002}{\doi{10.1103/PhysRevApplied.6.024002}}

\bibitem{Weichman2020}
P.~B. Weichman and E.~M. Lavely, ``\BIBforeignlanguage{en}{Fluorescent
  {{X}}-ray {{Scan Image Quality Prediction}}},''
  \emph{\BIBforeignlanguage{en}{J Hardw Syst Secur}}, vol.~4, no.~1, pp.
  13--23, Mar. 2020.
  \href{http://dx.doi.org/10.1007/s41635-019-00084-8}{\doi{10.1007/s41635-019-00084-8}}

\bibitem{Guruswamy2020}
T.~Guruswamy, L.~M. Gades, A.~Miceli, U.~M. Patel, J.~T. Weizeorick, and
  O.~Quaranta, ``Hard {{X}}-ray fluorescence measurements with {{TESs}} at the
  {{Advanced Photon Source}},'' in \emph{Journal of {{Physics}}: {{Conference
  Series}}}, vol. 1559.\hskip 1em plus 0.5em minus 0.4em\relax {IOP
  Publishing}, Sep. 2020, Art. no. 012018.
  \href{http://dx.doi.org/10.1088/1742-6596/1559/1/012018}{\doi{10.1088/1742-6596/1559/1/012018}}

\bibitem{Patel2020}
U.~Patel, R.~Divan, L.~Gades, T.~Guruswamy, D.~Yan, O.~Quaranta, and A.~Miceli,
  ``\BIBforeignlanguage{en}{Development of {{Transition}}-{{Edge Sensor X}}-ray
  {{Microcalorimeter Linear Array}} for {{Compton Scattering}} and {{Energy
  Dispersive Diffraction Imaging}}},'' \emph{\BIBforeignlanguage{en}{J. Low
  Temp. Phys.}}, vol. 199, no. 1-2, pp. 384--392, Apr. 2020.
  \href{http://dx.doi.org/10.1007/s10909-019-02267-7}{\doi{10.1007/s10909-019-02267-7}}

\bibitem{Ullom2004}
J.~N. Ullom, W.~B. Doriese, G.~C. Hilton, J.~A. Beall, S.~Deiker, W.~D. Duncan,
  L.~Ferreira, K.~D. Irwin, C.~D. Reintsema, and L.~R. Vale, ``Characterization
  and reduction of unexplained noise in superconducting transition-edge
  sensors,'' \emph{Appl. Phys. Lett.}, vol.~84, no.~21, pp. 4206--4208, May
  2004.  \href{http://dx.doi.org/10.1063/1.1753058}{\doi{10.1063/1.1753058}}

\bibitem{Brown2008}
A.-D. Brown \emph{et~al.}, ``\BIBforeignlanguage{en}{Absorber {{Materials}} for
  {{Transition}}-{{Edge Sensor X}}-ray {{Microcalorimeters}}},''
  \emph{\BIBforeignlanguage{en}{J Low Temp Phys}}, vol. 151, no.~1, pp.
  413--417, Apr. 2008.
  \href{http://dx.doi.org/10.1007/s10909-007-9669-2}{\doi{10.1007/s10909-007-9669-2}}

\bibitem{Yan2017a}
D.~Yan \emph{et~al.}, ``Eliminating the non-{{Gaussian}} spectral response of
  {{X}}-ray absorbers for transition-edge sensors,'' \emph{Appl. Phys. Lett.},
  vol. 111, no.~19, p. 192602, Nov. 2017.
  \href{http://dx.doi.org/10.1063/1.5001198}{\doi{10.1063/1.5001198}}

\bibitem{Mates2017}
J.~A.~B. Mates \emph{et~al.}, ``Simultaneous readout of 128 {{X}}-ray and
  gamma-ray transition-edge microcalorimeters using microwave {{SQUID}}
  multiplexing,'' \emph{Appl. Phys. Lett.}, vol. 111, no.~6, p. 062601, Aug.
  2017.  \href{http://dx.doi.org/10.1063/1.4986222}{\doi{10.1063/1.4986222}}

\bibitem{Mates2012}
J.~A.~B. Mates, K.~D. Irwin, L.~R. Vale, G.~C. Hilton, J.~Gao, and K.~W.
  Lehnert, ``Flux-ramp modulation for {{SQUID}} multiplexing,'' \emph{J. Low
  Temp. Phys.}, vol. 167, no. 5-6, pp. 707--712, Jun. 2012.
  \href{http://dx.doi.org/10.1007/s10909-012-0518-6}{\doi{10.1007/s10909-012-0518-6}}

\bibitem{Gard2018}
J.~D. Gard, D.~T. Becker, D.~A. Bennett, J.~W. Fowler, G.~C. Hilton, J.~A.
  Mates, C.~D. Reintsema, D.~R. Schmidt, D.~S. Swetz, and J.~N. Ullom, ``A
  scalable readout for microwave {{SQUID}} multiplexing of transition-edge
  sensors,'' \emph{J. Low Temp. Phys.}, vol. 193, no. 3-4, pp. 485--497, Nov.
  2018.
  \href{http://dx.doi.org/10.1007/s10909-018-2012-2}{\doi{10.1007/s10909-018-2012-2}}

\bibitem{Fowler2016}
J.~W. Fowler, B.~K. Alpert, W.~B. Doriese, Y.~I. Joe, G.~C. O'Neil, J.~N.
  Ullom, and D.~S. Swetz, ``The practice of pulse processing,'' \emph{J. Low
  Temp. Phys.}, vol. 184, no.~1, pp. 374--381, Jul. 2016.
  \href{http://dx.doi.org/10.1007/s10909-015-1380-0}{\doi{10.1007/s10909-015-1380-0}}

\bibitem{Levine1998a}
Z.~H. Levine and B.~Ravel, ``Identification of materials in integrated circuit
  interconnects using x-ray absorption near-edge spectroscopy,'' \emph{J. Appl.
  Phys.}, vol.~85, no.~1, pp. 558--564, Dec. 1998.
  \href{http://dx.doi.org/10.1063/1.369489}{\doi{10.1063/1.369489}}

\bibitem{Cressler2018}
J.~D. Cressler, \emph{\BIBforeignlanguage{en}{Fabrication of {{SiGe HBT BiCMOS
  Technology}}}}.\hskip 1em plus 0.5em minus 0.4em\relax {CRC Press}, Oct.
  2018.

\bibitem{Bohr1995}
M.~T. Bohr, ``Interconnect scaling-the real limiter to high performance
  {{ULSI}},'' in \emph{Proceedings of {{International Electron Devices
  Meeting}}}, Dec. 1995, pp. 241--244.
  \href{http://dx.doi.org/10.1109/IEDM.1995.499187}{\doi{10.1109/IEDM.1995.499187}}

\end{thebibliography}
 
\end{document}